\documentclass[showpacs,showkeys,prl,onecolumn]{revtex4}%
\usepackage{amsfonts}
\usepackage{amsmath}
\usepackage{amssymb}
\usepackage{graphicx}%
\begin{document}
\title{Effective interaction force between an electric charge and a magnetic dipole
and locality (or nonlocality) in quantum effects of the Aharonov-Bohm type.}
\author{$^{1}$Gianfranco Spavieri\footnote{gspavieri@gmail.com}, $^{2}$George T. Gillies, $^{3}$Miguel Rodriguez,
and $^{4}$Maribel Perez}
\affiliation{$^{1}$Centro de F\'{\i}sica Fundamental, Facultad de Ciencias, Universidad de
Los Andes, M\'{e}rida, 5101-Venezuela }
\affiliation{$^{2}$Department of Mechanical and Aerospace Engineering, University of
Virginia, Charlottesville, VA 22904 USA}
\affiliation{$^{3}$Carrera de F\'{\i}sica y Matem\'{a}tica, Facultad de Educaci\'{o}n, Universidad Nacional de Chimborazo, Riobamba,100150-Ecuador}
\affiliation{$^{4}$Carrera de Ingeniar\'{\i}a Ambiental, Facultad de Ingenier\'{\i}a, Universidad Nacional de Chimborazo, Riobamba,100150-Ecuador}
\email{gspavieri@gmail.com}

\begin{abstract}
Classical electrodynamics foresees that the effective interaction force
between a moving charge and a magnetic dipole is modified by the time-varying
total momentum of the interaction fields. We derive the equations of motion of
the particles from the total stress-energy tensor, assuming the validity of
Maxwell's equations and the total momentum conservation law. Applications to
the effects of Aharonov-Bohm type show that the observed phase shift may be
due to the relative lag between interfering particles caused by the effective
local force.

\end{abstract}

\maketitle

\textbf{1- Introduction. }

The discussions about the nonlocality, or locality, of the quantum effects of
the Aharonov-Bohm (AB) \cite{AB}-\cite{s2} type have been developing
extensively through the decades \cite{apv}-\cite{gc}. Some authors claim that
there are no forces acting on the particles in the AB effects \cite{apv},
\cite{lv}, \cite{bb}. However, other authors sustain diverse interpretations
of the AB type of effects in terms of the local action of forces \cite{boy},
\cite{boy1}, \cite{cv}, \cite{pl}, \cite{gc}, \cite{kk}. Moreover, it has been
argued that nonlocality claims are inconsistent with Hilbert-space quantum
mechanics \cite{rg}. Nonlocal quantum effects are relevant in a wide context
of scientific and heuristic scenarios and, according to the supporters of
quantum nonlocality, the effects of the AB type represent a new departure from
classical physics because they cannot be interpreted by means of local forces
acting on the particles. Traditional classical physics requires instead that
observable physical effects arise as a result of a cause, generally the action
of a force that produces the effect. Thus, supporters of classical physics
claim that the phase shift of the effects of the AB type may well arise as the
result of the action of local forces, which, for example, may produce a
relative lag between particles passing on opposite sides of the line sources,
resulting in an observable quantum phase shift $\Delta\phi$ \cite{boy},
\cite{boy1}, \cite{cv}, \cite{pl}, \cite{gc}, \cite{kk}.

As some other authors have pointed out \cite{bb}, we believe that there is no
consensus regarding the nature and interpretation of the AB effect, despite of
the several, ongoing discussions on the subject. Supporting the view of an
interpretation involving the local action of a force, we consider here three
effects of the AB type \cite{AB}-\cite{s2} where a beam of interfering
particles interacts with an external electromagnetic (em) potential (or field)
in a, supposedly, force-free region of space. In the AB effect \cite{AB} a
charged particle interacts with the vector potential $\mathbf{A}$ of a
solenoid, in the Aharonov-Casher (AC) effect \cite{AC} a neutral particle with
a magnetic dipole moment $\mathbf{m}$ interacts with an electric field
$\mathbf{E}$ of a line of charges, and in the Spavieri effect \cite{s2} a
neutral particle with an electric dipole moment $\mathbf{d}$ interacts with
the vector potential $\mathbf{A}$ of a distribution of magnetic dipoles. If
$\mathbf{Q}$ is the canonical interaction momentum, the quantum phase of these
effects is $\phi=\hbar^{-1}\int\mathbf{Q}\cdot d\vec{\ell}$. A unitary vision
of these effects is given in Ref. \cite{s2}, where $\mathbf{Q}$ is related to
the momentum of the interaction fields, $\mathbf{Q}=\pm\mathbf{Q}_{em}$. Some
of the phase shifts $\Delta\phi$ arising in these effects have been verified
experimentally: for the AB effect see Refs. \cite{cham}, \cite{tom} and, for
tests of the AC effect, see Refs. \cite{cim}, \cite{sang}.

The basic em interaction in the mentioned AB effects is the one between a
particle with an electric charge $q$ and a neutral particle possessing a
magnetic dipole moment $\mathbf{m}$. As well known, the standard expression
for the interaction force between a magnetic dipole $\mathbf{m}$ and a charge
$q$, in motion with relative velocity $\mathbf{v}$, does not comply with the
action and reaction principle. In fact, neglecting higher order relativistic
terms, in the reference frame where the dipole is stationary, the em force
acting on $q$ is $\mathbf{f}_{q}^{em}=q\mathbf{E}+(q/c)\mathbf{v\times
B}=(q/c)\mathbf{v\times B}$, where the electric field is $\mathbf{E}=0$ for a
neutral dipole and $\mathbf{B}$ is the magnetic induction field produced by
$\mathbf{m}$. Instead, the em force acting on $\mathbf{m}$ is $\mathbf{f}%
_{m}^{em}=\mathbf{\nabla(m\cdot B}_{q}\mathbf{)}$, where $\mathbf{B}%
_{q}=(\mathbf{v/}c)\mathbf{\times E}_{q}$ is the magnetic field produced by
the moving charge, $\mathbf{E}_{q}$ being its electric field. Thus, for
example, in the direction of motion, with $\mathbf{v}=v\widehat{i}$ in the $x$
direction, we find $f_{qx}^{em}=0$, while $f_{mx}^{em}=\mathbf{\partial}%
_{x}\mathbf{(m\cdot B}_{q}\mathbf{)}\neq0$. The action and reaction principle
is not conserved even when $q$ and $\mathbf{m}$ are stationary but the current
in $\mathbf{m}$ varies with time ($\mathbf{\dot{m}}\neq0)$, as pointed out by
Shockley-James \cite{sj}, who claim that their paradox indicates that even the
conservation law for the linear momentum is not conserved.

However, it has been shown (\cite{apv}, \cite{sn}) that the effective force
acting on a particle, and the corresponding equation of motion, might be
modified if, besides the mechanical momentum of the particle, (mass) $\times$
(velocity), the physical system carries an additional momentum related to the
interaction fields. In our case the fields of our physical system possesses a
nonvanishing electromagnetic interaction momentum $\mathbf{Q}_{em}$ and,
moreover, it possesses also a momentum $\mathbf{Q}_{h}$ that is due either to
the internal stresses \cite{apv}, \cite{sn} or to the charges induced by the
field $\mathbf{E}_{q}$ \cite{sn}. The purpose of our letter is to derive
within classical electrodynamics the equations of motion of $q$ and
$\mathbf{m}$ after determining the effective interaction force between them,
assuming the validity of conservation laws, the action and reaction principle,
and the contribution of the momentum of interaction fields, as also required
for the solution of the Shockley-James \cite{sj} paradox. The role of the
momentum $\mathbf{Q}_{h}$ has been taken into account in the nonrelativistic
interpretation of the atomic spin-orbit coupling \cite{gm} and we show that
the (effective) interaction force here derived leads to an interpretation of
the AB effects in terms of classical local forces. In fact, the derived
interaction force is in agreement with the experimental evidence of the
observed phase shift $\Delta\phi$ in the effects of the AB type \cite{cham}%
-\cite{sang}. Concerning the test of dispersionless forces through a long
toroid \cite{bb}, we show that our effective force agrees with the result of
the test. Other possible tests for the interaction force, realizable with
present technology \cite{gg}, are discussed.

\textbf{2- The interaction force between a charge }$q$\textbf{ and a magnetic
dipole moment }$\mathbf{m}$.

The relevant em interaction, taking place in the effects of the AB type, is
the one between a charged particle $q$ and a neutral particle with a magnetic
dipole moment $\mathbf{m}$. The appropriate tensor describing the system
composed by a charge $q$ and a nonconducting dipole $\mathbf{m}$ is
\cite{apv}, \cite{sn}, $T^{\mu\nu}=\theta^{\mu\nu}+S^{\mu\nu}+\delta_{0}%
U^{\mu}U^{\nu}$, complemented by the continuity equation, $\partial_{\mu
}T^{\mu\nu}=0$, where $\theta^{\mu\nu}$ is the em tensor, $S^{\mu\nu}$ is the
stress tensor, and $\delta_{0}$ is the proper density of the proper mass.
Relevant quantities are the em\textbf{\ }momentum $\mathbf{Q}_{em}$ and the
momentum $\mathbf{Q}_{h}$ due to stresses (referred to as the \textit{hidden}
momentum) \cite{apv}, \cite{sn}, their components being given by,%
\begin{equation}
Q_{em}^{i}=\frac{1}{4\pi c}\int(\mathbf{E\times B})^{i}\mathbf{\,}d\tau
=\int\theta^{i0}\mathbf{\,}d\tau;\quad Q_{h}^{i}=\int S^{i0}\mathbf{\,}d\tau.
\label{Ps}%
\end{equation}
Let $\mathbf{p}_{q}=M_{q}\mathbf{v}$ be the linear momentum of the charge of
mass $M_{q}$ and $\mathbf{p}_{m}=M_{m}\mathbf{v}_{m}$ that of the magnetic
dipole $\mathbf{m}$ of mass $M_{m}$. Integration over the volume of the
continuity equation, $\partial_{\mu}T^{\mu\nu}=\partial_{\mu}(\theta^{\mu\nu
}+S^{\mu\nu}+\delta_{0}U^{\mu}U^{\nu})$, leads to,%
\begin{equation}
\frac{d}{dt}(\mathbf{Q}_{em}+\mathbf{Q}_{h}\mathbf{)}+\int(\partial_{j}%
\theta^{ij}+\partial_{j}S^{ij})\mathbf{\,}d\tau+\frac{d}{dt}(\mathbf{p}%
_{q}+\mathbf{p}_{m}\mathbf{)}=0, \label{cm}%
\end{equation}

\begin{equation}
\frac{d}{dt}(\mathbf{Q}_{em}+\mathbf{Q}_{h}\mathbf{)}+\frac{d}{dt}%
(\mathbf{p}_{q}+\mathbf{p}_{m}\mathbf{)}=0,\label{eqm}%
\end{equation}
where, in (\ref{eqm}), for a closed isolated system the volume integral of the
divergences $\partial_{j}\theta^{ij}$ and $\partial_{j}S^{ij}$ vanishes, and
expression (\ref{cm}) provides the linear momentum conservation law
(\ref{eqm}). In the case of the interaction between $q$ and $\mathbf{m}$, in
the dipole approximation $\mathbf{Q}_{em}$ and $\mathbf{Q}_{h}$ can be
expressed \cite{boy}, \cite{hen}, \cite{apv}, \cite{sn}, as,%
\begin{equation}
\mathbf{Q}_{em}=\int\frac{\rho}{c}\mathbf{A}(\mathbf{x})d\tau=\frac{q}%
{c}\mathbf{A};\quad\mathbf{Q}_{h}=-\frac{q}{c}\mathbf{A}=\frac{\mathbf{m}}%
{c}\mathbf{\times E}_{q}.\label{pe}%
\end{equation}
Moreover, for a finite stationary configuration, expression (\ref{eqm})
implies \cite{apv}, \cite{sn},%

\begin{equation}
\mathbf{Q}_{em}+\mathbf{Q}_{h}=\frac{q}{c}\mathbf{A}+\frac{\mathbf{m}}%
{c}\mathbf{\times E}_{q}=0. \label{tp}%
\end{equation}
Since the interaction momentum $\mathbf{Q}_{h}$ is a nonlocal quantity, it can
be expressed indifferently in terms of $\mathbf{A}$ or $\mathbf{E}_{q}$ in
(\ref{pe}). In the dipole approximation $\mathbf{A}(\mathbf{x}-\mathbf{x}%
_{m})=\mathbf{m\times}(\mathbf{x}-\mathbf{x}_{m})/\left\vert \mathbf{x}%
-\mathbf{x}_{m}\right\vert ^{3}$ is the vector potential of the magnetic
dipole $\mathbf{m}$ and $\mathbf{E}_{q}=q(\mathbf{x}-\mathbf{x}_{q}%
)/\left\vert \mathbf{x}-\mathbf{x}_{m}\right\vert ^{3}$ is the electric field
of the charge $q$. In (\ref{pe}) $q\mathbf{A}$ has to be evaluated at the
position of the charge $\mathbf{x}_{q}$ and $c^{-1}\mathbf{m\times E}_{q}$ at
$\mathbf{x}_{m}$.

Making use of the relation $\partial_{\beta}\theta^{\alpha\beta}%
=-c^{-1}F^{\alpha\lambda}J\lambda$, where $F^{\alpha\lambda}$ is the em
field-strength tensor, after volume integration, for our closed system the em
and stress force density can be expressed respectively as, $\mathbf{f}%
^{em}=-(d/dt)\mathbf{Q}_{em}$ and $\mathbf{f}_{h}=-(d/dt)\mathbf{Q}_{h}$.
Then, $\mathbf{f}^{em}+\mathbf{f}_{h}=d(\mathbf{p}_{q}+\mathbf{p}%
_{m}\mathbf{)}/dt$\textbf{ }or,
\begin{equation}
\mathbf{f}^{em}-\frac{d}{dt}\mathbf{Q}_{h}=\frac{d}{dt}(\mathbf{p}%
_{q}+\mathbf{p}_{m}\mathbf{)=f}_{q}+\mathbf{f}_{m},\label{eqm1}%
\end{equation}
where in (\ref{eqm1}) $\mathbf{f}_{q}=d\mathbf{p}_{q}/dt$ and $\mathbf{f}%
_{m}=d\mathbf{p}_{m}/dt$ are the effective forces acting on $q$ and
$\mathbf{m}$ respectively. We can see from (\ref{eqm1}) that the effective
forces, $\mathbf{f}_{q}$ and $\mathbf{f}_{m}$, and the corresponding equations
of motion are affected by the term $d\mathbf{Q}_{h}/dt$. When the magnetic
dipole is made of conducting material, for the stationary system the electric
field $\mathbf{E}_{q}$ induces charges on the magnetic dipole creating an
electric field $\mathbf{E}_{ind}$ that provides zero total electric field
inside the dipole, $\mathbf{E}_{ind}+\mathbf{E}_{q}=0.$ In this case, if the
magnetic dipole is completely shielded, there are no internal stresses and no
momentum due to stresses related to $\mathbf{m}$ \cite{sn}. Still, the
original $\mathbf{Q}_{em}$ is modified now by the presence of the nonvanishing
$\mathbf{E}_{ind}$. In order to take into account the presence of
$\mathbf{E}_{ind}$, the relation $\mathbf{Q}_{em}+\mathbf{Q}_{h}=0$ of
(\ref{tp}) is replaced by the relation $\mathbf{Q}_{em}+\mathbf{Q}_{em-ind}=0$
where $\mathbf{Q}_{em-ind}=\mathbf{Q}_{h}$ is the em momentum related to the
induced charges and their field $\mathbf{E}_{ind}=-\mathbf{E}_{q}$ inside the
dipole. Thus, we assume here the validity of expression (\ref{tp}) with
$\mathbf{Q}_{h}$ representing, depending on the case, either the momentum due
to the stresses or the momentum due to the induced charges. The physical
results that can be derived from (\ref{eqm}) and (\ref{eqm1}) are supposed to
be model-independent and are shown to be holding for both the cases of
conducting and nonconducting magnetic dipole \cite{sn} in solving the
Shokley-James paradox \cite{sj}. Then, what (\ref{eqm1}) implies is that the
action of the time-variation of $\mathbf{Q}_{h}$ (or $\mathbf{Q}%
_{em-ind}=\mathbf{Q}_{h}$) has to be taken into account in (\ref{eqm1}) for
determining the effective $\mathbf{f}_{q}$ and $\mathbf{f}_{m}$ that comply
with the equilibrium condition and the momentum conservation law.

\textbf{3- Action and reaction principle. }

For our closed system we are left to split equation (\ref{eqm1}) into two
equations of motion, one for $q$ and one for $\mathbf{m}$. For our purposes,
we consider the case where $\mathbf{m}$ is stationary and $q$ is moving
relative to $\mathbf{m}$. The time derivative of $\mathbf{Q}_{em}$ can be
expressed as,%
\begin{align}
\mathbf{f}^{em}  &  =-\frac{d}{dt}\mathbf{Q}_{em}=\int(\rho\mathbf{E+j\times
B})\mathbf{\,}d\tau=q\mathbf{E}+\frac{q}{c}\mathbf{v\times B}+\int
(\mathbf{j\times B}_{q})\mathbf{\,}d\tau\label{eem}\\
&  =-\frac{q}{c}\partial_{t}\mathbf{A}+\frac{q}{c}\mathbf{v\times
B}+\mathbf{\nabla(m\cdot B}_{q}\mathbf{),}\nonumber
\end{align}
where in (\ref{eem}), $\mathbf{j\times B}=\mathbf{j}_{q}\mathbf{\times
B+j\times B}_{q}$. In (\ref{eem}) the terms $-\frac{q}{c}\partial
_{t}\mathbf{A}$ and $\frac{q}{c}\mathbf{v\times B}$ represent the standard
force on $q$ and the last term stands for the standard force \cite{jack},
\cite{boy}, \cite{boy1} on the magnetic dipole $\mathbf{m}$ in the presence of
the magnetic induction field $\mathbf{B}_{q}=c^{-1}\mathbf{v\times E}_{q}$ of
the moving charge. $\mathbf{Q}_{em}$ depends on the interacting fields of both
of $q$ and $\mathbf{m}$ and its time derivative (\ref{eem}) contains terms
representing forces localized and acting on $q$ as well as forces localized
and acting on $\mathbf{m}$. Being a nonlocal quantity, the momentum
$\mathbf{Q}_{em}$ is not localized on any of the field sources and thus its
time derivative, $-(d/dt)(\mathbf{Q}_{em}\mathbf{)}$, cannot be taken to
represent uniquely the forces acting on one of the interacting particles,
e.g., $q$ (or $\mathbf{m}$). Since similar considerations hold for
$\mathbf{Q}_{h}$, the variation $-(d/dt)(\mathbf{Q}_{h}\mathbf{)}$ represents
force terms, some acting on $q$ and others on $\mathbf{m}$ (and not uniquely
on $q$ or on $\mathbf{m}$).

When $\mathbf{j}_{q}=(\rho/c)\mathbf{v}$ forms part of a current loop in a
neutral wire, after integrating over the closed loop, the action and reaction
principle holds for interaction forces between the loop and the dipole
$\mathbf{m}$ even for time-varying fields. However, for our system the term
$(q/c)\mathbf{v}$ represents a non-neutral "open current" element and, due to
its nonvanishing electric field $\mathbf{E}_{q}$, the em momentum
$\mathbf{Q}_{em}\neq0$. As mentioned above, in this case we have
$(q/c)\mathbf{v\times B}+\mathbf{\nabla(m\cdot B}_{q}\mathbf{)}\neq0$, because
the force $(q/c)\mathbf{v\times B}$ is always perpendicular to the direction
of motion, while $\mathbf{\nabla(m\cdot B}_{q}\mathbf{)}$ has also a
nonvanishing longitudinal component in the direction of $\mathbf{v}$.
Moreover, the effective interaction force (between $q$ and $\mathbf{m}$) must
be such as to solve the Shokley-James paradox \cite{sj}, where the radiation
force on $q$, given by $-c^{-1}q\partial_{t}\mathbf{(A)}$, requires to be
balanced by an equal and opposite force on $\mathbf{m}$. As well known, this
paradox can be solved by taking into account the momentum $\mathbf{Q}_{h}$
\cite{apv}, \cite{sn} with the related force $\mathbf{f}_{h}\mathbf{\,}%
=-(d/dt)\mathbf{Q}_{h}$, which, on account of expression (\ref{pe}), may
conveniently written as,
\begin{align}
\mathbf{f}_{h}\mathbf{\,} &  =-\frac{d}{dt}\mathbf{Q}_{h}=-\frac{d}{dt}%
(\frac{\mathbf{m}}{c}\mathbf{\times\mathbf{E}}_{q})=\frac{d}{dt}(\frac{q}%
{c}\mathbf{A})\label{fd}\\
&  =-\frac{\mathbf{\dot{m}}}{c}\mathbf{\times\mathbf{E}}_{q}-\frac{\mathbf{m}%
}{c}\mathbf{\times(}\partial_{t}\mathbf{\mathbf{E}}_{q})=-\frac{\mathbf{\dot
{m}}}{c}\mathbf{\times\mathbf{E}}_{q}+\frac{q}{c}\mathbf{(v\cdot\nabla
)A.}\nonumber
\end{align}
When $\mathbf{v}=0$, the force $\mathbf{f}_{h}$ in (\ref{fd}) is
$-\partial_{t}\mathbf{(\mathbf{Q}}_{h})=$ $-c^{-1}\mathbf{(\dot{m}%
\times\mathbf{E}}_{q})=c^{-1}q\partial_{t}\mathbf{(A)}$, effective when the
current of the magnetic dipole is varying with time and em radiation fields
are involved. Then, the Shokley-James paradox is solved because the force
$-c^{-1}q\partial_{t}\mathbf{(A)}$, acting on $q$ in (\ref{eem}), is balanced
by the equal and opposite force $-c^{-1}\mathbf{(\dot{m}\times\mathbf{E}}%
_{q})=c^{-1}q\partial_{t}\mathbf{(A)}$ acting on $\mathbf{m}$.

Neglecting higher order relativistic terms, expression (\ref{tp}) holds even
when $q$ moves with velocity $\mathbf{v}$ relative to $\mathbf{m}$. Thus,
going beyond the effects of pure radiation fields, we assume that the idea
behind the Shokley-James paradox (that em interaction must comply with the
action and reaction principle and momentum conservation) can be extended to
include the velocity dependent terms of (\ref{eem}) and (\ref{fd}). For this
purpose it is sufficient that the equal and opposite interaction forces be
related to the gradients of the same interaction energy. The quantity
$-\mathbf{m}\cdot\mathbf{B}_{q}$ represents the interaction potential energy
between $\mathbf{m}$ and $\mathbf{B}_{q}$ and its negative $\mathbf{\nabla}$
leads to the force $\mathbf{\nabla(m\cdot B}_{q})$ acting on $\mathbf{m}$. In
order for the action and reaction principle to be holding, the same potential
energy must lead also to an equal and opposite force acting on $q$. In fact,
with the help of (\ref{pe}) and (\ref{tp}), we may write,%
\begin{equation}
-\mathbf{m}\cdot\mathbf{B}_{q}=-\mathbf{m}\cdot(\frac{\mathbf{v}}%
{c}\mathbf{\times\mathbf{E}}_{q})=\mathbf{v}\cdot(\frac{\mathbf{m}}%
{c}\mathbf{\times\mathbf{E}}_{q})=-\frac{q}{c}(\mathbf{v\cdot A}%
)\mathbf{,}\label{gr}%
\end{equation}
where $-(q/c)\mathbf{v\cdot A}$ represents the interaction potential energy
$-\int(\mathbf{j}_{q}\cdot\mathbf{A)\,}d\tau$ \cite{jack} of the open current
element $(q/c)\mathbf{v}$ in the presence of the vector potential $\mathbf{A}%
$. Then, expression (\ref{gr}) implies that, in correspondence to the force
term $\mathbf{\nabla(m}\cdot\mathbf{B}_{q})$ acting on the dipole, there is an
equal and opposite force $\mathbf{(}q/c)\mathbf{\nabla}(\mathbf{v\cdot
A})=\mathbf{(}q/c)\mathbf{v\times B}+\mathbf{(}q/c)\mathbf{(v\cdot\nabla)A}$
acting on the charge, as shown below.

\textbf{4- The equations of motion for the charge }$q$\textbf{ and the dipole
}$\mathbf{m}$.

As a criterion for identifying which are the effective forces on either $q$ or
$\mathbf{m}$, we assume the validity of the action and reaction principle.
With the help of Maxwell's equation $\mathbf{\nabla\times B}_{q}=\frac{1}%
{c}\partial_{t}\mathbf{E}_{q}$, equations (\ref{eem}), (\ref{fd}), and the
identity $\mathbf{\nabla(m\cdot B}_{q}\mathbf{)=(m\cdot\nabla})\mathbf{B}%
_{q}+\frac{1}{c}\mathbf{m\times}(\partial_{t}\mathbf{E}_{q})$, we may write
(\ref{eqm1}) as,%
\[
\mathbf{f}^{em}+\mathbf{f}_{h}=-\frac{q}{c}\partial_{t}\mathbf{A}+\frac{q}%
{c}\mathbf{v\times B}+\frac{q}{c}\mathbf{(v\cdot\nabla)A}%
\]%
\begin{equation}
-\frac{\mathbf{\dot{m}}}{c}\mathbf{\times\mathbf{E}}_{q}+\frac{\mathbf{m}}%
{c}\mathbf{\times(}\partial_{t}\mathbf{\mathbf{E}}_{q})+\mathbf{(m\cdot
\nabla)B}_{q}=\mathbf{f}_{q}+\mathbf{f}_{m}\label{qw}%
\end{equation}
The two terms of (\ref{qw}), $-\frac{q}{c}\partial_{t}\mathbf{A}$ and
$-\frac{\mathbf{\dot{m}}}{c}\mathbf{\times E}$, are equal and opposite and,
algebraically, may cancel. However, from a physical point of view, we may not
suppress them if they represent, as they do in this context, the equal and
opposite action and reaction forces on $q$ and $\mathbf{m}$, respectively.
About the terms $\frac{q}{c}\mathbf{(v\cdot\nabla)A}$ and $\frac{\mathbf{m}%
}{c}\mathbf{\times(}\partial_{t}\mathbf{\mathbf{E}}_{q})$ of (\ref{qw}) (which
are equal and opposite and, algebraically, may cancel) we consider the
following possible interpretations.

\textit{a}) As\ done by Aharonov et al. \cite{apv}, we may assume that the
momentum $\mathbf{Q}_{h}$ is localized on the dipole $\mathbf{m}$. Then, in
this case the force (\ref{fd}) $-(d/dt)\mathbf{Q}_{h}=-\frac{\mathbf{\dot{m}}%
}{c}\mathbf{\times E}+\frac{q}{c}\mathbf{(v\cdot\nabla)A}=-\frac
{\mathbf{\dot{m}}}{c}\mathbf{\times E-}\frac{\mathbf{m}}{c}\mathbf{\times
(}\partial_{t}\mathbf{\mathbf{E}}_{q})$ is entirely acting on $\mathbf{m}$ and
$\frac{q}{c}\mathbf{(v\cdot\nabla)A}$ and $\frac{\mathbf{m}}{c}\mathbf{\times
(}\partial_{t}\mathbf{\mathbf{E}}_{q})$ in (\ref{qw}) are equal and opposite
force terms that cancel because both act on $\mathbf{m}$. This way, the
longitudinal components of the forces disappear and expression (\ref{qw})
becomes,%
\begin{equation}
\mathbf{f}_{q}+\mathbf{f}_{m}=-\frac{q}{c}\partial_{t}\mathbf{A}+\frac{q}%
{c}\mathbf{v\times B-}\frac{\mathbf{\dot{m}}}{c}\times\mathbf{\mathbf{E}%
+\mathbf{(m\cdot\nabla)B}}_{q},
\end{equation}
where the first two terms on the rhs represent $\mathbf{f}_{q}$ and the last
two represent $\mathbf{f}_{m}$.

\textit{b}) Nevertheless, as mentioned above, the momentum is a nonlocal
quantity and, as such, theoretically and experimentally, neither
$\mathbf{Q}_{em}$ or $\mathbf{Q}_{h}$ can be localized on any of the source of
the interaction fields. What can be assumed as localized are the forces that
are related to the time (or space) variations of $\mathbf{Q}_{em}$ or
$\mathbf{Q}_{h}$. Therefore, a priori, we cannot exclude that the term
$\frac{q}{c}\mathbf{(v\cdot\nabla)A}$ may represent a force acting on the
current element $(q/c)\mathbf{v}$. In this case, we may not cancel
algebraically the two terms $\frac{q}{c}\mathbf{(v\cdot\nabla)A}$ and
$\frac{\mathbf{m}}{c}\mathbf{\times(}\partial_{t}\mathbf{\mathbf{E}}_{q})$ in
(\ref{qw}) because they represent, respectively, a force acting on $q$ and a
force on $\mathbf{m}$. Thus, being $\mathbf{v\times B}+\mathbf{(v\cdot
\nabla)A}=\mathbf{\nabla}(\mathbf{v\cdot A})$, in the reference frame where
the magnetic dipole is stationary, expression (\ref{qw}) leads to the
following effective forces and corresponding equations of motion,%
\begin{equation}
\mathbf{f}_{q}=-\frac{q}{c}\partial_{t}\mathbf{A}+\frac{q}{c}\mathbf{\nabla
}(\mathbf{v\cdot A})=\frac{d}{dt}(\mathbf{p}_{q}\mathbf{)} \label{efq}%
\end{equation}

\begin{equation}
\mathbf{f}_{m}=-\frac{\mathbf{\dot{m}}}{c}\mathbf{\times E}+\mathbf{\nabla
(m\cdot B}_{q})=\frac{d}{dt}(\mathbf{p}_{m}\mathbf{).} \label{efm}%
\end{equation}
The Lagrangian formulation for deriving (\ref{efq}) and (\ref{efm}) will be
given in a future contribution.

\textbf{5- Experimental evidence. }

The correct choice, \textit{a}) or \textit{b}), for the effective force on the
charge $q$ in motion, needs to be corroborated experimentally. In the
experiment by Becker and Batelaan \cite{bb} the long macroscopic toroid,
adopted for testing the time of flight of $q$, has not been used to observe
the AB effect. Moreover, it is reasonable to assume that the vector potential
$\mathbf{A}$ is nearly uniform inside the long toroid where, in the direction
of motion, $\mathbf{(v\cdot\nabla)A}\simeq0$. Thus, no action on the moving
charge is exerted inside the toroid by the force term $\frac{q}{c}%
\mathbf{(v\cdot\nabla)A}$ of (\ref{efq}) and, consequently, no variation of
its time of flight is foreseen, in agreement with the experimental results
\cite{bb}. In fact, we expect our force to be acting briefly on the moving $q$
at the beginning and end of the toroid only. The ideal experiment to more
easily detect the longitudinal component of the force must adopt an
arrangement where the perpendicular component $\mathbf{v\times B}$ vanishes
and, moreover, $\mathbf{A}$ is not uniform in the direction of motion, so that
$\mathbf{(v\cdot\nabla)A}_{\shortparallel}\neq0$. Such an arrangement is
obtained in the case of the AB effect with a standard toroid, where the force
(\ref{efq}) is nonvanishing. Actually, in the experiment performed by Tomonura
et al. \cite{tom}, which detects the Aharonov-Bohm phase shift $\Delta
\phi_{AB}$, a microscopic toroid has been used and the result corroborates the
existence of a longitudinal force.

The effect of a local force on the phase shift of the AB system has been
derived in Refs. \cite{boy}, \cite{boy1} and we reconsider it here starting
from the free-force phase $\phi=\mathbf{\hbar}^{-1}(\mathbf{p\cdot x}-\bar
{E}t)$ of the interfering wave function of the electrons in the AB effect. In
the presence of the small force $\mathbf{f}_{q}$, the particles moving on
opposite sides of the solenoid acquire a relative lag that produces the phase
shift $\Delta\phi$, either because of the particles relative variation
$\delta\mathbf{v}$ or $\delta\mathbf{x}$ \cite{boy}, \cite{boy1}, \cite{sn}.
Assuming that the particle of mass $M_{q}$ slightly changes its momentum
$\mathbf{p}_{q}=M_{q}\mathbf{v}$ under the action of the force $\mathbf{f}%
_{q}=M_{q}d\mathbf{v}/dt$, we have $\delta\mathbf{v}=M_{q}^{-1}\int_{0}%
^{t}\mathbf{f}_{q}dt$. If $x(t)$ is the position of the particle when
$\mathbf{p}_{q}$ is constant, with $vdt=dx$ along the path of the particle
(from $-\infty$ to $+\infty$), the force has the effect to change it by
$\Delta x=\int(\delta v)dt=M_{q}^{-1}\int dt\int_{0}^{t}(\mathbf{f}_{q}%
)_{x}dt=(vM_{q})^{-1}\int dt\int_{0}^{t}[\mathbf{\nabla}(\frac{q}%
{c}\mathbf{A\cdot v})]_{x}dx$. Then, $\Delta x=(vM_{q})^{-1}\int_{-\infty
}^{\infty}\frac{q}{c}\mathbf{A\cdot v}dt$, leading to the phase shift
$\delta\phi=\mathbf{\hbar}^{-1}(\mathbf{p}_{q}\cdot\Delta\mathbf{x}%
)=\mathbf{\hbar}^{-1}\frac{q}{c}\int_{-\infty}^{\infty}\mathbf{A}\cdot
d\mathbf{x}$. Because of the topological properties of the system and the
symmetry of vector potential $\mathbf{A}$, the resulting relative phase shift
between particles moving along the opposite sides of the solenoid is,%
\begin{equation}
\Delta\phi=-2\mathbf{\hbar}^{-1}\frac{q}{c}\int_{-\infty}^{\infty}%
\mathbf{A}\cdot d\mathbf{x=\hbar}^{-1}\frac{q}{c}%
{\textstyle\oint_{C}}
\mathbf{A}\cdot d\mathbf{x}=\Delta\phi_{AB},\label{fab}%
\end{equation}
where $\Delta\phi_{AB}$ is the observable Aharonov-Bohm phase shift.

Similar conclusions may be drawn for the Aharonov-Casher \cite{AC} and
Spavieri \cite{s2} effects. In the case of the AC effect, $\mathbf{m}$ is
moving with velocity $\mathbf{v}_{m}=-\mathbf{v}$ relative to a static
electric charge distribution. Then, in (\ref{efm}) the term $\mathbf{\nabla
(m\cdot B})=-c^{-1}\mathbf{\nabla(m\cdot v}_{m}\times\mathbf{E}%
)=\mathbf{\nabla(\mathbf{P}}_{m}\cdot\mathbf{E})=(\mathbf{P}_{m}%
\cdot\mathbf{\nabla)E}$ represents the force on the electric dipole moment
$\mathbf{P}_{m}=c^{-1}\mathbf{v}_{m}\times\mathbf{m}$ of the moving
$\mathbf{m}$. Thus, the force on the moving magnetic dipole $\mathbf{m}$ turns
out to be given by the same expression as derived by Boyer in his classical
interpretation of the AC effect \cite{boy}, \cite{boy1}. The experimental
evidence for the existence of the local force is given by the experiments
cited in \cite{cim} and \cite{sang}.

\textbf{6- Concluding remarks}

We have derived within classical electrodynamics the expressions (\ref{efq})
and (\ref{efm}) for the effective interacting force on the charge $q$ and on
the dipole $\mathbf{m}$, respectively. Starting from first principles, we
describe the isolated system by means of the total tensor $T^{\mu\nu}$
assuming the validity of the continuity equation $\partial_{\mu}T^{\mu\nu}=0$
and the conservation law of the total linear momentum. Our effective force
expressions, which solve the Shockley-James paradox, account for a force term
acting in the direction of motion. This term foresees that an observable phase
shift must take place between interfering particles encircling the magnetic
flux in the AB effect. The phase shift occurs because of the relative lag of
the particles produced by the local action of the force. The best experimental
evidence for the existence of the longitudinal force term in the AB and AC
effects, is given by the observed phase shift in the tests of Refs.
\cite{cham}-\cite{sang}.

The tradition in physics, based on cause and effect, requires observed effects
to be explained by means of the local causes that produce them, be their
origin due to electromagnetic interaction or even interacting quantum systems.
With our approach, the observed AB phase shifts can be interpreted in terms of
the action of local em effective forces, reinforcing the view of the classical
origin of the effects of the AB type. For a conclusive interpretation of the
Aharonov-Bohm effect it is essential, as suggested in Ref. \cite{bb}, to close
the loopholes  that exist in the tests of the em interaction. Ideally, tests
of the em forces acting in the $q-\mathbf{m}$ interaction (some are discussed
in Ref. \ref{gg}) should aim at verifying the validity of the action and
reaction principle, the conservation laws, and the conclusive expressions of
both forces $\mathbf{f}_{q}$ and $\mathbf{f}_{m}$.

\textbf{7- Acknowledgments}

We thank the CDCHT (ULA, M\'{e}rida, Venezuela) for partial support.

\end{document}